\documentclass[11pt, a4paper,english,superscriptaddress,aps]{revtex4}
\usepackage{amsmath,amssymb,graphicx}
\makeatletter
\usepackage{babel}
\usepackage[active]{srcltx}
\usepackage{graphicx,color}
\usepackage{changebar}
\usepackage{hyperref}
\usepackage[T1]{fontenc}
\usepackage{ae}
\usepackage[ansinew]{inputenc}
\usepackage{amsmath}
\usepackage{braket}
\usepackage{mathtools}
\usepackage{slashed}
\usepackage{empheq}

\usepackage{graphicx}
\usepackage{amsfonts}
\usepackage{amsmath}
\newcommand{\be}{\begin{equation}}
\newcommand{\ee}{\end{equation}}
\newcommand{\bea}{\begin{eqnarray}}
\newcommand{\eea}{\end{eqnarray}}

\begin{document}

\title{Traversable wormholes in Chern-Simons modified gravity}

\author{J. R. Nascimento}
\email[]{jroberto@fisica.ufpb.br}
\affiliation{Departamento de F\'{\i}sica, Universidade Federal da 
Para\'{\i}ba\\
 Caixa Postal 5008, 58051-970, Jo\~ao Pessoa, Para\'{\i}ba, Brazil}

\author{A. Yu. Petrov}
\email[]{petrov@fisica.ufpb.br}
\affiliation{Departamento de F\'{\i}sica, Universidade Federal da 
Para\'{\i}ba\\
 Caixa Postal 5008, 58051-970, Jo\~ao Pessoa, Para\'{\i}ba, Brazil}

\author{P. J. Porf\'{i}rio}\email[]{pporfirio89@gmail.com}
\affiliation{Departamento de F\'{\i}sica, Universidade Federal da Para\'{\i}ba\\
 Caixa Postal 5008, 58051-970, Jo\~ao Pessoa, Para\'{\i}ba, Brazil}

\begin{abstract}
In this paper, we examine the existence of traversable wormhole solutions within the Chern-Simons modified gravity. We find a non-trivial solution in the  theory with dynamical Chern-Simons coefficient in the absence of matter sources. This result displays a situation opposite to GR where the  matter sources violating the energy conditions are required.
\end{abstract}

\maketitle

\section{Introduction}

In recent years, the search for modified gravity theories as an alternative to General Relativity has strongly intensified.  The main reasons for this consist in the facts that, first, the GR does not allow for a consistent quantum description of gravity due to its non-renormalizability, second, it fails to explain the cosmic acceleration. In this context, various alternative gravity models are considered, such as $f(R)$ gravity, $f(R,Q)$ gravity, Horava-Lifshitz gravity models, theories including torsion and nonmetricity and many other examples. Among these models, an important role is played by the four-dimensional Chern-Simons (CS) modified gravity \cite{JaPi}, {whose action looks like sum of the four-dimensional gravitational  CS term with the usual GR action, so that it involves a Chern-Pontryagin density (third derivatives of the metric background) coupled to a pseudoscalar field $\vartheta$ called the CS coefficient}.  In particular, this theory represents itself as a first known example of CPT-breaking (and, for a special form of the CS coefficient, Lorentz-breaking) gravity model. Additionally, the CS modified gravity also turns out to be relevant in other physical contexts such as: string theory (after dimensional reduction process) \cite{Polch}, loop quantum gravity and particle physics, see f.e. \cite{AlYunes} and references therein. Further extension of CS modified gravity called the dynamical CS modified gravity included a nontrivial dynamics for $\vartheta$ as well (for a general review on CS modified gravity, both in dynamical and non-dynamical cases, see f.e. \cite{AlYunes}). 

As it is known, consistency of any modified gravity theory is verified through checking of consistency of known metric solutions in this theory. In \cite{Grumiller},  a wide class of spherically and axially symmetric solutions of equations of motion in CS modified gravity was analyzed. In \cite{Pretorius}, it was shown that to achieve order-by-order perturbative consistency of the Kerr metric within this theory, it should be modified by the $\vartheta$-dependent terms, and in \cite{ours}, the consistency of G\"{o}del-type solutions in this theory was verified, with the causality conditions were analyzed. 

Among other interesting gravity solutions, the wormholes are of a special importance. Their key feature consists in the fact that in certain cases they allow for trajectories connecting two causally disconnected points. Another remarkable feature about the wormholes is the possibility of time travel, namely, it is possible for an observer in such geometries to travel faster than light. It is worth pointing out that around each point of this geometry the local Lorentz invariance is fulfilled, but globally this is not hold. It was shown in \cite{Lobo} the procedure how to build a time machine from wormholes. Many interesting results related to wormholes can be found in \cite{Lobo,KJ}. However, up to now, the consistency of wormhole solutions within the CS modified gravity never was studied. This is the aim of this paper.

The structure of this work looks like follows. In the section 2, we give a general description of geometry of traversable wormholes. In the section 3, we present a brief review on CS modified gravity. And in the section 4, we verify the consistency of wormholes within the dynamical CS modified gravity and discuss the related energy conditions. Finally, in the section 5 we present our conclusions. In the Appendix A we present an explicit form of components of the Cotton tensor and the energy-momentum tensor of the $\vartheta$ field.

\section{Traversable Wormhole geometries}

Wormholes have initially been introduced by Einstein and Rosen in seminal paper \cite{EinsteinRosen}. This wormhole (Einstein-Rosen bridge) has constructed from a particular extension of maximally extended Schwarzschild solution of General Relativity. It covers two asymptotically flat spaces connected by a ``bridge" whose spatial section possesses a non-trivial topology. Another important feature is that Einstein-Rosen wormhole presents event horizons, thus an observer traveling along wormhole could not cross it, in other words, it is forbidden traveling from one flat space to the other. However, there must be conditions which prevent the emergence of the event horizons. Such conditions are so-called of traversability conditions and the wormholes that satisfied such conditions are known by traversable wormholes.

 In this section we briefly discuss the most important features of the traversable wormholes. Its line element in spherical coordinates, following  \cite{Morris}, is given by
\begin{equation}
ds^2=-e^{2\Phi(r)}dt^2+ \frac{dr^2}{1-b(r)/r}+r^{2}(d\theta^2+\sin^{2}\theta d\phi^2).
\label{wormhole}
\end{equation}
Note that this line element has spherical symmetry and is asymptotically flat. Furthermore, the radial coordinate $r$ is non-monotonic, it may be split into two patches: the first one covers the range where $r$ decreases from $+\infty$ to $r_{0}$, where $r_{0}$ (represents the throat of the wormhole) is the minimum of $r$, and the second one covers the range where $r$ increases from $r_{0}$ to $+\infty$.  
The functions $\Phi(r)$ and $b(r)$ stands for the gravitational redshift and shape function of the wormhole, respectively. In general, both functions must fulfill some requirements (traversability conditions): in order to avoid the presence of horizons it is necessary that $e^{\Phi(r)}$ is positive everywhere or, identically, $\Phi(r)$ must be finite and essentially real. Concerning the shape function we can show using embedding arguments: it must satisfy the flaring-out condition at the throat, i.e.,
\begin{equation}
\frac{(b-b^{\prime}r)}{2b^2}>0,	
\end{equation}
where the prime means derivative with respect $r$, leads to $b^{\prime}(r_{0})<1$ since $b(r_0)=r_0$, as a consequence the proper distance will be minimal at the throat \cite{Morris}. 

In GR, such a traversable wormhole geometries requires matter sources violating the null energy condition (NEC), see f.e. \cite{Morris,Wheeler,Lobo}. This violation is represented by the inequality
\begin{equation}
T^{(m)}_{\mu\nu}k^{\mu}k^{\nu}<0,
\end{equation}
 where $k^{\mu}$ represents any null vector and $T^{(m)}_{\mu\nu}$ is the energy-momentum tensor of the matter sources. Indeed, the NEC is obtained from the famous Raychaudhuri equation whose form for null geodesics looks like  
\begin{equation}
\frac{d\theta}{d\tau}=-\frac{1}{2}\theta^{2}-\sigma_{\mu\nu}\sigma^{\mu\nu}+\omega_{\mu\nu}\omega^{\mu\nu}-R_{\mu\nu}k^{\mu}k^{\nu},
\label{Ray}
\end{equation}
where $\theta$ is the expansion, $\sigma_{\mu\nu}$ denotes the shear, $\omega_{\mu\nu}$ represents the vorticity all them related to the null geodesics congruence  defined by the tangent null vector $k_{\mu}$ that emerges the traversable wormhole at one side and going out at the other one. As shown in \cite{Visser,Lobo}, the vorticity is identically zero for a traversable wormhole. In addition, at the throat the expansion $\theta$ is also zero, satisfying the condition $\dfrac{d\theta}{d\tau}\geq 0$, thus Eq. (\ref{Ray}) reduces to inequality
\begin{equation}
R_{\mu\nu}k^{\mu}k^{\mu}\leq 0,
\label{geo}
\end{equation}
implying that null geodesics will defocus so that the null convergence condition \cite{HawkingEllis} does not hold. By means of the Einstein field equations, the direct consequence of Eq. (\ref{geo}) can be shown to look like
\begin{equation}
T^{(m)}_{\mu\nu}k^{\mu}k^{\nu}\leq0,
\end{equation}
therefore, the NEC is either violated or on the imminent of being violated \cite{VisserPaper}.
We note that actually our model represents itself as a gravity coupled to a scalar (either ghost or not). The coupling of the usual general relativity to the scalar matter has been studied in great details in \cite{Fisher} for a non-ghost matter, and in \cite{BerLei} for a ghost matter. As we see further, our theory involves additional non-minimal coupling responsible for the Chern-Simons term. It was argued in \cite{Bron1,HEllis} that these solutions display a wormhole-like behavior.

\section{Chern-Simons modified gravity}

In the current section we present a brief review of the four-dimensional Chern-Simons (CS) modified gravity. Its action can be cast into the form \cite{Jackiw,Yunes}
\begin{equation}
 S_{CS}=\frac{1}{2\kappa}\int d^{4}x \sqrt{-g}R+\frac{\alpha}{4}\int d^{4}x \sqrt{-g}\bigg(\vartheta\ ^{*}RR\bigg)-\frac{\tilde{\beta}}{2}\int d^{4}x\sqrt{-g}\bigg(\partial^{\mu}\vartheta\partial_{\mu}\vartheta\bigg)+S_{mat}, 
\label{ActionCS} 
 \end{equation}
where $\kappa=8\pi G$,  $\vartheta$ is a (pseudo)-scalar field, $\alpha$ and $\tilde{\beta}$ are coupling constants with a non-zero mass dimension, and $S_{mat}$ is the matter action. Let us make substitutions $\alpha=1/2\kappa$ and $\tilde{\beta}=\beta/\kappa$, thus the dimension of $\vartheta$ is length squared, $\left[\vartheta\right]=L^2$, whereas $\beta$ has the dimension of inverse fourth power of length, $\left[\beta\right]=L^{-4}$. Regarding to second term in the action, the $\ ^{*}RR$ is a topological term (it does not contribute for the equations of motion, if $\vartheta=const$) called the Chern-Pontryagin term and defined by
\begin{equation}
 \begin{split}
 \ ^{*}RR &\equiv \ ^{*}R^{\mu\ \ \gamma\sigma}_{\ \ \nu}R^{\nu}_{\ \mu\gamma\sigma}=\frac{1}{2}\frac{\epsilon^{\gamma\sigma\tau\eta}}{\sqrt{-g}}R^{\mu}_{\ \ \nu\tau\eta}R^{\nu}_{\ \mu\gamma\sigma},\\
 \end{split}
 \end{equation}
 where $\epsilon^{\gamma\sigma\tau\eta}$ is the Levi-Civita symbol and $\ ^{*}R^{\mu\ \ \gamma\sigma}_{\ \ \nu}$ is the dual Riemann tensor. In more mathematical language, the Chern-Pontryagin term in Eq.(\ref{ActionCS}) can be rewritten as a second order polynomial of the two-form curvature, $Tr(R\wedge R)$ \cite{ChernSimons}, such a quantity is proportional to the well-known second Chern class which, in turn, is related to Chern-Simons form $\Omega$ by means of a total derivative \cite{Eguchi},
\begin{equation}
Tr(R\wedge R)=d\Omega,
\end{equation}
  where $\Omega=Tr\bigg(\omega\wedge d\omega+\dfrac{2}{3}\omega\wedge\omega\wedge\omega\bigg)$ and $\omega$ is one-form connection. In a more physical notation, the former equation becomes 
		\begin{equation}
		\ ^{*}RR=2\,\nabla_{\mu}K^{\mu},
		\end{equation}
   where $\Gamma^{\lambda}_{\nu\sigma}$ represents the Levi-Civita connection coefficients, and $K^{\mu}=\dfrac{\epsilon^{\mu\nu\alpha\beta}}{\sqrt{-g}}\bigg(\Gamma^{\lambda}_{\nu\sigma}\partial_{\alpha}\Gamma^{\sigma}_{\beta\lambda}+\frac{2}{3}\Gamma^{\lambda}_{\nu\sigma}\Gamma^{\sigma}_{\alpha\theta}\Gamma^{\theta}_{\beta\lambda}\bigg)$ is the topological current. Therefore, by integrating by parts the second term in Eq. (\ref{ActionCS}) we get
			\begin{equation}
			\int d^{4}x \sqrt{-g}\bigg(\vartheta\ ^{*}RR\bigg)=-2\int d^{4}x \sqrt{-g}\,v_{\mu}K^{\mu},
			\label{Lorentz}
			\end{equation}
where $v_{\mu}=\partial_{\mu}\vartheta$. 
When the $\vartheta$ becomes a constant, the modified theory reduces to the usual General Relativity.
	
	The modified field equations are obtained varying to the action with respect to the metric and scalar field.  Doing this, we arrive at:
	\begin{eqnarray}
G_{\mu\nu}+ C_{\mu\nu}&=&\kappa T_{\mu\nu}^{(m)}+T^{(\vartheta)}_{\mu\nu};\nonumber\\
\beta\square\vartheta&=&-\frac{1}{4}\,^{*}RR,
\label{KG}
\end{eqnarray}
where  $\square\equiv\dfrac{1}{\sqrt{-g}}\partial_{\mu}(\sqrt{-g}g^{\mu\nu}\partial_{\nu})$ is for the covariant d'Alembertian operator. The energy-momentum tensor is defined as follows: first, the
\begin{equation}
T_{\mu\nu}^{(m)}=-\frac{2}{\sqrt{-g}}\bigg(\frac{\delta\mathcal{L}_{m}}{\delta g^{\mu\nu}} \bigg),
\end{equation}
describes the energy-momentum tensor of the matter sources and
\begin{equation}
T^{(\vartheta)}_{\mu\nu}=\beta\left[(\partial_{\mu}\vartheta) (\partial_{\nu}\vartheta)-\frac{1}{2}g_{\mu\nu}(\partial_{\lambda}\vartheta) (\partial^{\lambda}\vartheta)\right],
\end{equation}
represents the energy-momentum tensor of the contributions of $\vartheta$. Finally, the variation of the second term in Eq. (\ref{ActionCS}) with respect to the metric gives rise to the Cotton tensor $C^{\mu\nu}$ explicitly written as
\begin{eqnarray}
 C^{\mu\nu}=-\frac{1}{2}\bigg[v_{\sigma}\bigg(\frac{\epsilon^{\sigma\mu\alpha\beta}}{\sqrt{-g}}\nabla_{\alpha}R^{\nu}_{\beta}+
\frac{\epsilon^{\sigma\nu\alpha\beta}}{\sqrt{-g}}\nabla_{\alpha}R^{\mu}_{\beta}\bigg)+v_{\sigma\tau}({}^{*}R^{\tau\mu\sigma\nu}+\ ^{*}R^{\tau\nu\sigma\mu})\bigg],
\end{eqnarray}    
 where 
 $v_{\sigma\tau}=\nabla_{\sigma}v_{\tau}$. 

The theory may be considered within two approaches depending on $\beta$ coupling, namely: the first one is for non-dynamical framework (non-dynamical CS, NCS theory) that implies $\beta=0$ so that the kinetic term is ruled out, the second one is for the dynamical framework (dynamical CS, DCS theory), in this case $\beta\neq 0$ involves the non-zero kinetic term.  

\section{Traversable Wormholes in CS modified gravity}

\subsection{Vacuum solution}

In this section we examine the possibility of the existence of traversable wormhole vacuum solutions both in non-dynamical and dynamical frameworks.

To study the equations of motion in our theory, for convenience, we will use the Cartan formalism.  Within its methodology, for the traversable wormhole manifolds given by (\ref{wormhole}) we can define a local Lorentz (orthonormal) co-frame such that
\begin{eqnarray}
\theta^{(0)}&=& e^{\Phi(r)}dt;\nonumber\\
\theta^{(1)}&=& (1-b(r)/r)^{-1/2}dr;\nonumber\\
\theta^{(2)}&=&r d\theta;\nonumber\\
\theta^{(3)}&=&r\sin\, \theta d\varphi,
\label{TB}
\end{eqnarray}
where $ds^2=\eta_{AB}\theta^{A}\theta^{B}$, with $\eta_{AB}=diag(-1,+1,+1,+1)$ being the Minkowski metric. It was shown in \cite{JaPi} that in the Schwarzschild case one has $^{*}RR=0$, and in \cite{Yunes}, the same result was shown to hold  for all spherically symmetric geometries. Taking into account this result, the field equations in the co-frame (\ref{TB}), in the absence of matter,  take the form
\begin{eqnarray}
\label{32} G_{AB}+C_{AB}&=&T^{(\vartheta)}_{AB},\\
\label{33} \beta\square \vartheta&=&0,
\end{eqnarray} 
where we have used natural units, $\kappa=1$.
The Ricci tensor and the Einstein tensor in the co-frame (\ref{TB}) are diagonal. Furthermore, for the sake of simplicity, we will take $\Phi^{\prime}=0$ which leads to a constant redshift function (zero tidal force); such a condition avoids possible higher derivative terms, as well as other complications in the field equations. Carrying out these simplifications, we have
\begin{equation}
\begin{split}
G_{(0)(0)}&=\frac{b^{\prime}(r)}{r^2};\\
G_{(1)(1)}&=-\frac{b(r)}{r^3};\\
G_{(2)(2)}&=G_{(3)(3)}=-\frac{1}{2}\bigg(\frac{b^{\prime}(r)}{r^2}-\frac{b(r)}{r^3}\bigg).
\end{split}
\end{equation}

The non-vanishing components of $C_{AB}$ and $T^{(\vartheta)}_{AB}$ are described in Appendix A. 

\subsection{Non-Dynamical framework}

This case is covered by taking $\beta=0$ in the field equations implying that $T^{(\vartheta)}_{AB}$ is ruled out. Hence, the Eqs. (\ref{32},\ref{33}) reduce to
\begin{eqnarray}
\label{34} G_{AB}+C_{AB}&=&0.
\end{eqnarray}
Note that the former equation decouples for spherically symmetric metrics as shown in \cite{Grumiller}, namely, it can be rewritten as follows
\begin{eqnarray}
R_{AB}&=&0,\\
\label{Birk}C_{AB}&=&0,
\end{eqnarray}
where the first equation is identical to the vacuum Einstein equation.

As a result of Birkhoff theorem, the field equations (\ref{Birk}) do not describe wormholes. Hence, we conclude that the CS modified gravity in the absence of matter, in non-dynamical framework, does not support wormhole-like geometries independently of the CS field form.

\subsection{Dynamical framework}

Differently from the non-dynamical case, the dynamical framework ($\beta\neq 0$) is a more rich approach because the CS field is treated as a dynamical one. As a consequence, the field equations may dramatically change allowing wormhole vacuum solutions in contrast to NCS theory. In fact, it is reasonable to expect that, because in this case, the energy-momentum tensor of usual matter is zero, but the dynamics of CS scalar field adds up a new contribution to the right-hand side of usual Einstein equations, so, the space-time now possesses an extra amount of energy (scalar hair).

 Before proceeding with the components of Eq. (\ref{32}), let us focus on Eq. (\ref{33}). It represents itself as a massless Klein-Gordon equation, and in the spaces (\ref{wormhole}) it can be solved via separation of variables. More precisely, within the classification used in \cite{Ellis1, Ellis2, Ellis3} such geometries describe locally rotational space-time (LRS), and their Killing vectors yield the Lie algebra $T_{1}\oplus so(3)$, where $T_{1}$ represents the Killing vector $\partial_{t}$ due to the fact that Eq. (\ref{wormhole}) is static while $so(3)$ is associated to other three Killing vectors with one of them being $\partial_{\phi}$, implying rotational symmetry. Adding all this to the fact of linearity of (\ref{32}), we conclude that it can be solved through  separation of variables. As shown in \cite{Canfora}, the invariant (maximally symmetric) solutions of Eq. (\ref{33}) have the form $\vartheta=\vartheta(x^{l})e^{\gamma x^{j}}$, where $j$ labels the coordinates of the symmetries of (\ref{wormhole}), and $l$ labels other coordinates.

In particular, as the CS coefficient is a real pseudo-scalar field, it can be written as 
\begin{equation}
\label{sol} \vartheta(t,r,\theta,\phi)=A(r)Y(\theta)\textbf{Re}\bigg(e^{i(m\phi+\omega t)}\bigg).
\end{equation}
 Substituting it in Klein-Gordon equation, we get
\begin{eqnarray}
\label{angular}\bigg[\frac{1}{\sin(\theta)}\frac{d}{d\theta}\bigg(\sin(\theta)\frac{d}{d\theta}\bigg)-\bigg(\frac{m}{\sin(\theta)}\bigg)^{2}\bigg]Y(\theta)&=&-l(l+1) Y(\theta),\\
\label{radial}(r^{2}-b r)A^{\prime\prime}-\bigg(\frac{3}{2}b-2r +\frac{1}{2}r b^{\prime}\bigg)A^{\prime}+r^{2}\omega^{2}A&=&l(l+1) A,
\end{eqnarray}
 where the prime stands for derivative with respect to $r$ and $l(l+1)$ is the separation constant ($l=0,1,2...$). The Eq. (\ref{angular}) is nothing more as the well known associated Legendre equation whose solutions are given by associated Legendre  polynomials. On the other hand, the latter equation imposes the relation between $A(r)$ and $b(r)$.   

Returning to the modified field equations (\ref{32}), we may simplify the solution that comes from Klein-Gordon equation. For this, one must note that the off-diagonal field equations decouple, and, then, the field equations reduce to a set of partial differential equations (PDEs) which must be solved for non-trivial $\vartheta$.
Accordingly, we have
\begin{eqnarray}
\label{G} G_{AB}&=&T^{(\vartheta)}_{AB},\quad \mbox{A=B}\\
\label{Ndia} C_{AB}&=&T^{(\vartheta)}_{AB},\quad \mbox{A$\neq$ B}\\
\label{box} \square\vartheta &=& 0.
\end{eqnarray}
It is clear that wormhole solutions in DCS theory are also solutions in GR by requiring the vanishing of Eq. (\ref{Ndia}) for some non-trivial $\vartheta$ satisfying the other equations, (\ref{G}) and (\ref{box}).
 
Now, we shall solve the system of PDEs (\ref{Ndia}). As a first step, let us substitute the CS coefficient of the form (\ref{sol}) into the non-vanishing components of this system, explicitly discriminated in Appendix A.
We get that the non-trivial solutions must fulfill the followings requirements: first, the separation constant should be zero ($l=0$). As a consequence one has $m=0$ as well, second, $\omega=0$. As a result of that, it arrives that the only solution of Eq. (\ref{angular}) is $\frac{d Y(\theta)}{d\theta}=0$. Therefore, the field equations themselves force the CS coefficient to take the form $\vartheta(t,r,\theta,\phi)=A(r)$ leading to the vanishing of all the components of Cotton tensor and non-diagonal components of $T_{AB}^{(\vartheta)}$. 

We may exactly solve the radial equation that reduces to
\begin{equation}
(r^{2}-b r)A^{\prime\prime}-\bigg(\frac{3}{2}b-2r +\frac{1}{2}r b^{\prime}\bigg)A^{\prime}=0,
\end{equation}     
whose solution can be expressed in terms of the shape function, namely, 
\begin{equation}
A(r)=\gamma\bigg[\int \frac{1}{r^2}\bigg(1-\frac{b(r)}{r}\bigg)^{-1/2} dr\bigg],
\label{eqA}
\end{equation} 
where $\gamma$ is an integration constant whose dimension is length cubed. The gradient of $A(r)$ is actually the vector $v_{\mu}$ defined in previous section. Notice that $v_{A}=\bigg[0\, ,\frac{\gamma}{r^2},\,0,\,0\bigg]$, evaluated in Lorentz co-frame (\ref{TB}), represents a space-like vector with a preferred direction on space-time. Therefore, there is a preferred local Lorentz-frame where $v_{A}$ takes a non-zero value. It is interesting to note  the fact that $v_{A}$ does not depends on $b(r)$. 


Following this procedure, remains us to solve the diagonal components of the modified field equations. Note that the CS field in the form (\ref{eqA}) gives rise to simplest non-trivial components of energy-momentum tensor of CS field, obtained in the co-frame (\ref{TB}) in which they do not depend on $b(r)$:  
\begin{equation}
\begin{split}
T^{(\vartheta)}_{(0)(0)}&=\frac{\beta}{2}\frac{\gamma^2}{r^4};\\
T^{(\vartheta)}_{(1)(1)}&=\frac{\beta}{2}\frac{\gamma^2}{r^4};\\
T^{(\vartheta)}_{(2)(2)}&=-\frac{\beta}{2}\frac{\gamma^2}{r^4};\\
T^{(\vartheta)}_{(3)(3)}&=-\frac{\beta}{2}\frac{\gamma^2}{r^4}.
\label{TEM}
\end{split}
\end{equation}
In addition, the CS field in the form (\ref{eqA}) leads to the vanishing Cotton tensor.

Inserting them in the modified field equations, we have a solution that leads to a specific form for the shape function,
\begin{equation}
b(r)=-\frac{1}{2}\frac{\beta \gamma^{2}}{r},
\label{eqb}
\end{equation}
explicitly depending upon $\gamma$ and $\beta$ parameters. In particular, the shape function given by Eq. (\ref{eqb}) only obeys the flaring-out condition when $\beta$ is strictly negative. However, such a choice on $\beta$ leads to the wrong-sign kinetic term in the action (\ref{ActionCS}) so that the energy of the theory is not bounded from below, in other words, perturbations around the vacuum expectation value are unstable. 
Effectively, our manner to introduce the dynamics of the Chern-Simons coefficient $\vartheta$ provides a ghost-like dynamics for it (for discussion of different issues related to ghosts, see f.e. \cite{Hawking}). Also, we note that the use of the exotic matter is necessary for existence of wormholes and other noncausal solutions, see f.e. \cite{Morris,Lobo}. Nevertheless, we should emphasize that the $\vartheta$ is not a common matter field but an ingredient of the CS modified gravity, so, this theory itself contains the possibility of wormhole solutions, even in a vacuum, since there is no common matter contributions in (\ref{Ndia}). In order to meet the minimum distance condition at the throat, it is necessary which $r_{0}^{2}=-\dfrac{1}{2}\beta \gamma^{2}$ implying 
\begin{equation}
b(r)=\frac{r_{0}^{2}}{r}.
\end{equation}
 It just putting the above equation into Eq. (\ref{eqA}) one finds an explicit form of CS field, namely,
\begin{equation}
\vartheta(r)=A(r)=\frac{\pi}{2r_0}-\frac{1}{r_0}\arctan\bigg(\sqrt{\frac{r^2}{r_{0}^{2}}-1}\bigg).
\label{eqvar}
\end{equation}
So, our main result consists in a possibility to find the shape function $b(r)$ from the known CS coefficient $\vartheta(r)$, or vice versa.

\begin{figure}
	\centering
				\includegraphics[scale=0.8]{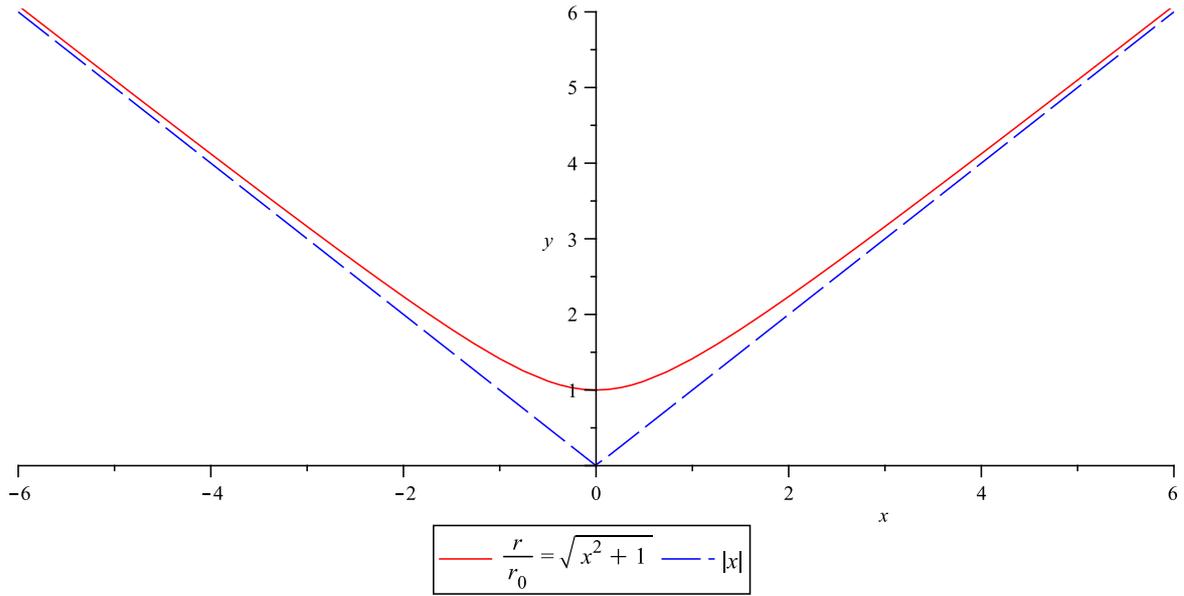}
		\caption{\label{CSgraph2} The figure displays $r/r_{0}$ in terms of $x$. The dashed lines show the behavior asymptotic corresponding to Minkoswkian space that implies $b(r)\rightarrow 0$ and $\vartheta(r)\rightarrow 0$.}
\end{figure}
\begin{figure}\centering
	\includegraphics[scale=0.6]{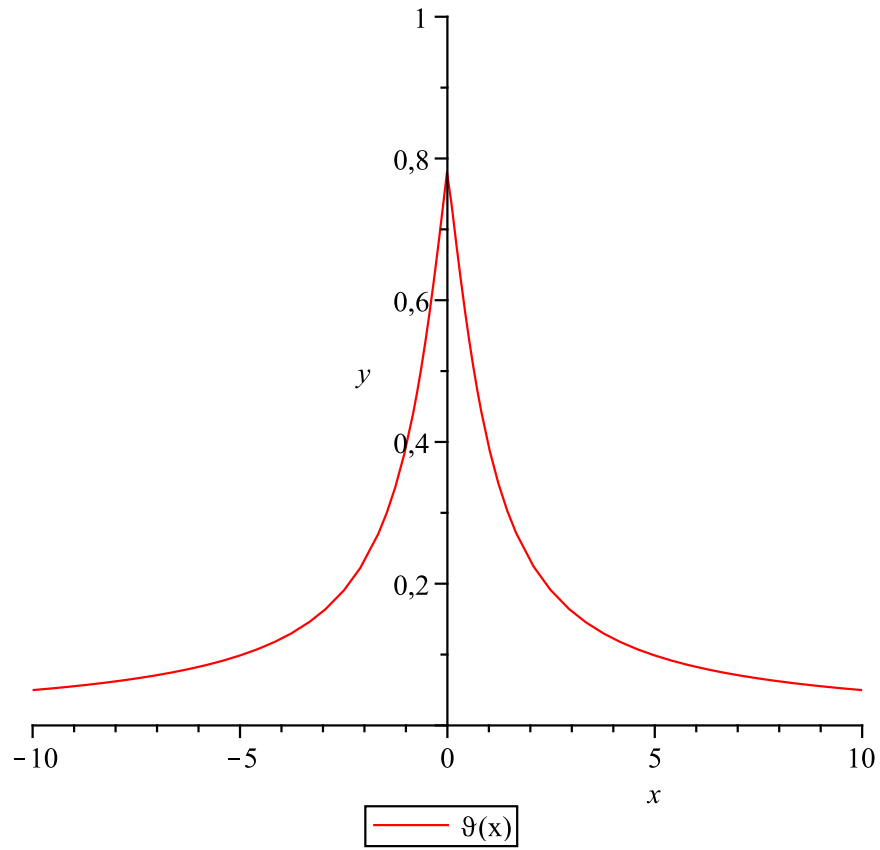}
\caption{\label{CSgraph} The graph shows the behavior of CS field versus $x$. We take $r_{0}=2$.}
\end{figure}

It is worth calling attention that the $\arctan\bigg(\sqrt{\frac{r^2}{r_{0}^{2}}-1}\bigg)$ is a multi-valued function then, taking this into account and the fact that its argument is strictly positive, its image is inside the interval $[0,\pi/2)$. The solution takes a maximal value at the throat, $\vartheta(r_0)=A(r_0)=\frac{\pi}{2r_0}$, and goes to zero far away from it, i.e.,
\begin{equation}
\lim_{r\rightarrow\pm\infty} \frac{1}{r_0}\arctan\bigg(\frac{1}{\sqrt{\frac{r^2}{r_{0}^{2}}-1}}\bigg)=0.
\end{equation}
For $\frac{r}{r_0}-1\ll 1$, we can approximate $\arctan\bigg(\sqrt{\frac{r^2}{r_{0}^{2}}-1}\bigg)\simeq \sqrt{2(\frac{r}{r_{0}}-1)}$.
 By making such a choice, the solution does not respect the $\vartheta$ shift symmetry (the constant cannot be dropped), so there one meets the spontaneous breaking down of the shift symmetry. On the other hand, the solution holds the symmetry of the metric under transformation $r\rightarrow-r$. For a more detailed analysis, we shall make the following transformation:
\begin{equation}
 |x|=\sqrt{\frac{r^2}{r_{0}^{2}}-1},
\label{tra}
\end{equation}
where the modulus has been used to ensure the symmetry under transformation $x\rightarrow-x$. Note that the minimal value of $|x|$  corresponds to Eq. ({\ref{tra}}) evaluated at the throat, $x(r_{0})=0$. This transformation is convenient because it introduces a monotonic variable (one chart) covering the whole wormhole instead of two charts. All the aforementioned features are displayed in Figs. (\ref{CSgraph2},\ref{CSgraph}) as well as the global behavior of $x$ and $\vartheta(r)$.  
%

 
The energy-momentum tensor of CS field given by Eq. (\ref{TEM}) may be reinterpreted through its splitting into two parts: the first contribution, $T_{AB}^{(d)}$, is identical to a dust-like  contribution with negative energy density, thus violating the null energy condition  (which is not unusual since the wormhole solutions require exotic matter, see f.e. \cite{Lobo}), while the second contribution, $T_{AB}^{(e)}$ has a form analogous to a source-free circularly symmetric electric or magnetic field, where the explicit form of $T_{AB}^{(e)}$ is given by \cite{Olmo2016}. Thus, we can relate the radius of the throat $r_0$ with an effective electric charge $q$ through the expression $r_{0}^{2}=\dfrac{q^2}{8\pi}$; this charge arises from a flux due to a non-trivial topology of our background. Explicitly,  we have 
\begin{equation}
T_{AB}^{(\vartheta)}=T_{AB}^{(d)}+T_{AB}^{(e)},
\end{equation}
where 
\begin{equation}
T_{AB}^{(d)}=-\dfrac{2r_{0}^{2}}{r^4}\,\mbox{diag}(1,0,0,0),
\end{equation}
 and 
\begin{equation}
T_{AB}^{(e)}=\dfrac{r_{0}^{2}}{r^4}\,\mbox{diag}(1,-1,1,1).
\end{equation}
So, the CS field in the presence of the background metric (\ref{wormhole}) behaves as a combination of the dust-like matter violating the energy conditions and a free electric charge (as expected such a behavior is identical to that one in GR, see \cite{Sushkov}). Such a interpretation arises due to the non-trivial topology of (\ref{wormhole}); in \cite{Wheeler1}, this mechanism is referred as charge without charge.



\subsection{Non-null redshift}

As argued in \cite{Yunes}, the condition $^{*}RR=0$ does not necessarily imply in vanishing of the Cotton tensor, for an explicit example see \cite{ours}. Having this in mind, a natural generalization of the metric (\ref{wormhole}) is the case characterized by a non-constant redshift, $g_{00}\neq const$. Our aim in this subsection is to inspect the influence of the non-constant redshift on the Cotton tensor. For this purpose, let us use a more general wormhole metric, see f.e. \cite{Bronnikov},
\begin{equation}
ds^2=-A(x)dt^2+\frac{1}{B(x)}dx^2+r(x)^2(d\theta^2+\sin^{2}\theta d\phi^2),
\label{Gwormhole}
\end{equation}
where $A(x)$, $B(x)$ and $r(x)$ are functions of the arbitrary radial coordinate $x$. Similarly to Eq.(\ref{wormhole}), the metric \ref{Gwormhole} must satisfy some extra conditions for describing a traversable wormhole geometry, namely, $r(x)$ has a global minimum at a point $x=x_{0}$, as mentioned in the Section $1$ such a point is referred by throat, further $A(x)$ must be positive definite in order to prevent event horizons around the throat. For sake of simplicity, but without loss of generality, it is natural to make the gauge choice, $A(x)=B(x)$.

We shall take a similar ansatz to (\ref{sol}) for the CS field, i.e., $\vartheta(t,x,\theta,\phi)=U(x)Y(\theta)$. The non-vanishing components of the Einstein, Cotton and energy-momentum tensor in the coordinate basis are explicitly displayed in Appendix B. Thus, the field equations again decouple and take the form
\begin{eqnarray}
\label{DAG} G_{\nu}^{\mu}&=&T_{\phantom{(\vartheta)}\nu}^{(\vartheta)\,\mu},\quad \nu=\mu\\
C_{\nu}^{\mu}&=&T_{\phantom{(\vartheta)}\nu}^{(\vartheta)\,\mu}, \quad \nu\neq\mu\\
\square\vartheta&=& 0.
\end{eqnarray}

By subtracting the $2,2$ and $3,3$ components of Eq. (\ref{DAG}) we get 
\begin{equation}
0=U(x)\frac{d}{d\theta}Y(\theta) \Rightarrow \frac{d}{d\theta}Y(\theta)=0,
\end{equation}
imposing the constraint that non-trivial solutions have the form $\vartheta(t,x,\theta,\phi)=U(x)$ culminating in the cancellation of the Cotton tensor and off-diagonal components of $T^{(\vartheta)\,\mu}_{\phantom{(\vartheta)}\nu}$.

Briefly, we have shown in both cases: with and without constant redshift, that the field equations naturally decouple for wormhole-like geometries (\ref{wormhole},\ref{Gwormhole}). The non-trivial solutions require that the CS field has an exclusive dependence on the radial coordinate. On the other hand, such a requirement leads to the vanishing of Cotton tensor. We conclude that even starting from DCS theory, the field equations reduce to GR and scalar field ones, and our results are in accordance with those ones suggested in \cite{Grumiller} for the dynamical framework.


 \section{Summary}

We verified the conditions of persistence of wormhole-like vacuum solutions within CS modified gravity. The vacuum solutions in non-dynamical framework are identical to GR ones. Therefore, wormhole geometries are not vacuum solutions of NCS theory.

Unlike non-dynamical framework,  we have shown that wormhole vacuum solutions are allowed in dynamical framework. We realized that the off-diagonal modified field equations in local Lorentz frame lead to a non-trivial setup for CS field $\vartheta$, satisfying its evolution equation. Such a field configuration implies in vanishing of the Cotton tensor as well as the off-diagonal energy-momentum tensor of $\vartheta$.

We found the CS field setup allowing for the wormhole solution. Besides, the flaring-out condition presumes $\beta$ to be essentially negative, as a consequence the kinetic energy associated to the scalar field has a wrong sign, thus it behaves like a ghost field. It turns out to be that this field may be seen like a combination of ghost-like dust and electromagnetic field, moreover, the non-trivial properties of the wormhole geometry lead us to a topological charge associated to the throat of the wormhole. This solution setup reduces to the same as GR plus scalar field, as it can be expected because the Cotton tensor disappears, whilst the diagonal components of energy-momentum tensor of $\vartheta$ continue to be non-zero.

By considering non-zero tidal force we have arrived at the same conclusions as in the case of zero redshift, i.e., the modified field equations for DCS theory in vacuum always will reduce to GR and scalar field ones, independently of the redshift function.
The exotic character of the CS field is nevertheless natural -- indeed, it is well known that noncausal solutions like wormholes, Alcubierre warp drive etc., require exotic matter, see f.e. \cite{Lobo}.
The natural generalization of this study could consist in considering a generic situation where a Cotton tensor does not vanish. We plan to carry out this generalization in a forthcoming paper.

\section{Appendix A. Non-vanishing components of the Cotton and energy-momentum tensors}

In this appendix we write the non-zero components of the Cotton and energy-momentum tensor. For this calculus, we used the {\it GRtensor} program. Accordingly, for the Cotton tensor we get
\begin{eqnarray}
\nonumber C_{(0)(2)}&=&-\frac{1}{4\,r^{5}\,\sin(\theta)}\bigg(1-\frac{b(r)}{r}\bigg)^{-1/2}\, \bigg[ 3\,b \left( r
 \right) {\frac {\partial }{\partial \phi}}\vartheta \left( t,r,\theta,\phi
 \right) -3\,b \left( r \right) r{\frac {\partial ^{2}}{\partial r
\partial \phi}}\vartheta \left( t,r,\theta,\phi \right)-\\
\nonumber &-&3\, \left( {\frac 
{\partial }{\partial \phi}}\vartheta \left( t,r,\theta,\phi \right) 
 \right) r b^{\prime} \left( r \right)+{r}^{2} \left( {\frac {
\partial ^{2}}{\partial r\partial \phi}}\vartheta \left( t,r,\theta,\phi
 \right)  \right) b^{\prime} \left( r \right)+\\
 &+& \left( {\frac {
\partial }{\partial \phi}}\vartheta \left( t,r,\theta,\phi \right) 
 \right) {r}^{2} b^{\prime\prime} \left( r \right)\bigg];\\
\nonumber C_{(0)(3)}&=&-\frac{1}{4\,r^{5}}\bigg(1-\frac{b(r)}{r}\bigg)^{-1/2}\, \bigg[ 3\,b \left( r
 \right) {\frac {\partial }{\partial \theta}}\vartheta \left( t,r,\theta,\phi
 \right) -3\,b \left( r \right) r{\frac {\partial ^{2}}{\partial r
\partial \theta}}\vartheta \left( t,r,\theta,\phi \right)-\\
\nonumber &-&3\, \left( {\frac 
{\partial }{\partial \theta}}\vartheta \left( t,r,\theta,\phi \right) 
 \right) r b^{\prime} \left( r \right)+{r}^{2} \left( {\frac {
\partial ^{2}}{\partial r\partial \theta}}\vartheta \left( t,r,\theta,\phi
 \right)  \right) b^{\prime} \left( r \right)+\\
 &+& \left( {\frac {
\partial }{\partial \theta}}\vartheta \left( t,r,\theta,\phi \right) 
 \right) {r}^{2} b^{\prime\prime} \left( r \right)\bigg];\\
C_{(1)(2)}&=&-\frac{1}{4 {r}^{4}\sin \left( \theta
 \right) }\,{ \left( {\frac 
{\partial ^{2}}{\partial t\partial \phi}}\vartheta \left( t,r,\theta,\phi
 \right)  \right)  \left( 3\,b \left( r \right) -r b^{\prime}
 \left( r \right)  \right) };\\
C_{(1)(3)}&=&\frac{1}{4\,r^4}\,{ \left( {\frac 
{\partial ^{2}}{\partial t\partial \theta}}\vartheta \left( t,r,\theta,\phi
 \right)  \right)  \left( 3\,b \left( r \right) -r b^{\prime}
 \left( r \right)  \right) }.
\end{eqnarray}

The non-vanishing components of $T_{AB}^{\vartheta}$ are
\begin{eqnarray}
\nonumber T_{(0)(0)}^{\vartheta}&=&\frac{1}{2\,{r}^{2} \left( \sin \left( \theta \right)  \right) ^{2}}\,\bigg[- \left( {\frac {
\partial }{\partial r}}\vartheta \left( t,r,\theta,\phi \right)  \right) ^{
2}r \left( \sin \left( \theta \right)  \right) ^{2}b \left( r \right) 
+ \left( {\frac {\partial }{\partial \phi}}\vartheta \left( t,r,\theta,\phi
 \right)  \right) ^{2}+\\
\nonumber &+& \left( {\frac {\partial }{\partial t}}\vartheta
 \left( t,r,\theta,\phi \right)  \right) ^{2}{r}^{2} \left( \sin
 \left( \theta \right)  \right) ^{2}+ \left( {\frac {\partial }{
\partial \theta}}\vartheta \left( t,r,\theta,\phi \right)  \right) ^{2}
 \left( \sin \left( \theta \right)  \right) ^{2}+\\
&+& \left( {\frac {
\partial }{\partial r}}\vartheta \left( t,r,\theta,\phi \right)  \right) ^{
2}{r}^{2} \left( \sin \left( \theta \right)  \right) ^{2}\bigg];\\
T_{(0)(1)}^{\vartheta}&=&-\bigg(1-\frac{b(r)}{r}\bigg)^{-1/2} \left( {\frac {
\partial }{\partial t}}\vartheta \left( t,r,\theta,\phi \right)  \right) {
\frac {\partial }{\partial r}}\vartheta \left( t,r,\theta,\phi \right);\\
T_{(0)(2)}^{\vartheta}&=&-\frac{1}{r} \left( {
\frac {\partial }{\partial t}}\vartheta \left( t,r,\theta,\phi \right) 
 \right) {\frac {\partial }{\partial \theta}}\vartheta \left( t,r,\theta,
\phi \right);\\
T_{(0)(3)}^{\vartheta}&=&-\frac{1}{r\sin \left( \theta \right) }\left( {
\frac {\partial }{\partial t}}\vartheta \left( t,r,\theta,\phi \right) 
 \right) {\frac {\partial }{\partial \phi}}\vartheta \left( t,r,\theta,\phi
 \right);\\
\nonumber T_{(1)(1)}^{\vartheta}&=&-\frac{1}{2{r}^{2} \left( 
\sin \left( \theta \right)  \right) ^{2}}\bigg[ \left( {\frac {
\partial }{\partial r}}\vartheta \left( t,r,\theta,\phi \right)  \right) ^{
2}r \left( \sin \left( \theta \right)  \right) ^{2}b \left( r \right) 
- \left( {\frac {\partial }{\partial r}}\vartheta \left( t,r,\theta,\phi
 \right)  \right) ^{2}{r}^{2} \left( \sin \left( \theta \right) 
 \right) ^{2}-\\
\nonumber &-& \left( {\frac {\partial }{\partial t}}\vartheta \left( t,r,
\theta,\phi \right)  \right) ^{2}{r}^{2} \left( \sin \left( \theta
 \right)  \right) ^{2}+ \left( {\frac {\partial }{\partial \theta}}
\vartheta \left( t,r,\theta,\phi \right)  \right) ^{2} \left( \sin \left( 
\theta \right)  \right) ^{2}+\\
&+& \left( {\frac {\partial }{\partial \phi}
}\vartheta \left( t,r,\theta,\phi \right)  \right) ^{2}\bigg];\\
T_{(1)(2)}^{\vartheta}&=&\frac{1}{r} \bigg(1-\frac{b(r)}{r}\bigg)^{-1/2} \left( {\frac {
\partial }{\partial \theta}}\vartheta \left( t,r,\theta,\phi \right)  \right) {
\frac {\partial }{\partial r}}\vartheta \left( t,r,\theta,\phi \right);\\
T_{(1)(3)}^{\vartheta}&=&\frac{1}{r\sin \left( \theta \right) }\bigg(1-\frac{b(r)}{r}\bigg)^{-1/2}\left( {\frac {
\partial }{\partial r}}\vartheta \left( t,r,\theta,\phi \right)  \right) {
\frac {\partial }{\partial \phi}}\vartheta \left( t,r,\theta,\phi \right);\\
\nonumber T_{(2)(2)}^{\vartheta}&=&-\frac{1}{2{r}^{2} \left( 
\sin \left( \theta \right)  \right) ^{2}}\bigg[-\left( {\frac {
\partial }{\partial r}}\vartheta \left( t,r,\theta,\phi \right)  \right) ^{
2}r \left( \sin \left( \theta \right)  \right) ^{2}b \left( r \right) 
- \left( {\frac {\partial }{\partial \theta}}\vartheta \left( t,r,\theta,
\phi \right)  \right) ^{2} \left( \sin \left( \theta \right)  \right) 
^{2}+\\
\nonumber &+& \left( {\frac {\partial }{\partial r}}\vartheta \left( t,r,\theta,
\phi \right)  \right) ^{2}{r}^{2} \left( \sin \left( \theta \right) 
 \right) ^{2}- \left( {\frac {\partial }{\partial t}}\vartheta \left( t,r,
\theta,\phi \right)  \right) ^{2}{r}^{2} \left( \sin \left( \theta
 \right)  \right) ^{2}+\\
&+& \left( {\frac {\partial }{\partial \phi}}\vartheta
 \left( t,r,\theta,\phi \right)  \right) ^{2}\bigg];\\
T_{(2)(3)}^{\vartheta}&=&\frac{1}{{r}^{2}\sin \left( \theta \right) } \left( {
\frac {\partial }{\partial \theta}}\vartheta \left( t,r,\theta,\phi
 \right)  \right) {\frac {\partial }{\partial \phi}}\vartheta \left( t,r,
\theta,\phi \right);\\
\nonumber T_{(3)(3)}^{\vartheta}&=&\frac{1}{2{r}^{2} \left( 
\sin \left( \theta \right)  \right) ^{2}}\bigg[\left( {\frac {
\partial }{\partial r}}\vartheta \left( t,r,\theta,\phi \right)  \right) ^{
2}r \left( \sin \left( \theta \right)  \right) ^{2}b \left( r \right) 
+ \left( {\frac {\partial }{\partial \phi}}\vartheta \left( t,r,\theta,\phi
 \right)  \right) ^{2}-\\
\nonumber &-& \left( {\frac {\partial }{\partial r}}\vartheta
 \left( t,r,\theta,\phi \right)  \right) ^{2}{r}^{2} \left( \sin
 \left( \theta \right)  \right) ^{2}+ \left( {\frac {\partial }{
\partial t}}\vartheta \left( t,r,\theta,\phi \right)  \right) ^{2}{r}^{2}
 \left( \sin \left( \theta \right)  \right) ^{2}-\\
&-& \left( {\frac {
\partial }{\partial \theta}}\vartheta \left( t,r,\theta,\phi \right) 
 \right) ^{2} \left( \sin \left( \theta \right)  \right) ^{2}\bigg]
\end{eqnarray}

\section{Appendix B. Non-vanishing components of the Einstein, Cotton and energy-momentum tensors for the metric \ref{Gwormhole}}

The non-vanishing components of $G^{\mu}_{\nu}$ are
\begin{eqnarray}
G^{0}_{0}&=&{\frac {2\,A \left( x \right) r \left( x \right) {\frac {d^{2}}{d{x}^{
2}}}r \left( x \right) + \left( {\frac {d}{dx}}A \left( x \right) 
 \right) r \left( x \right) {\frac {d}{dx}}r \left( x \right) -1+A
 \left( x \right)  \left( {\frac {d}{dx}}r \left( x \right)  \right) ^
{2}}{ \left( r \left( x \right)  \right) ^{2}}}
 ,\\
G^{1}_{1}&=&{\frac { \left( {\frac {d}{dx}}A \left( x \right)  \right) r \left( x
 \right) {\frac {d}{dx}}r \left( x \right) -1+A \left( x \right) 
 \left( {\frac {d}{dx}}r \left( x \right)  \right) ^{2}}{ \left( r
 \left( x \right)  \right) ^{2}}}
,\\
G^{2}_{2}&=& G^{3}_{3}= \frac{1}{2}\,\Bigg[{\frac {2\, \left( {\frac {d}{dx}}A \left( x \right)  \right) {
\frac {d}{dx}}r \left( x \right) +2\,A \left( x \right) {\frac {d^{2}}
{d{x}^{2}}}r \left( x \right) + \left( {\frac {d^{2}}{d{x}^{2}}}A
 \left( x \right)  \right) r \left( x \right) }{r \left( x \right) }}\Bigg].
\end{eqnarray}

The non-vanishing components of $C^{\mu}_{\nu}$ are
\begin{eqnarray}
\nonumber C^{0}_{3}&=&\frac {1}{4\left( r \left( x
 \right)  \right) ^{2}}\Bigg[\sin \left( \theta \right)\left( {\frac {d}{d\theta}}Y
 \left( \theta \right)  \right)\Bigg]  \Bigg( U \left( x \right)  \left( r
 \left( x \right)  \right) ^{2}{\frac {d^{3}}{d{x}^{3}}}A \left( x
 \right) -\nonumber\\ &-&
 4\,U \left( x \right)  \left( {\frac {d}{dx}}A \left( x
 \right)  \right) r \left( x \right) {\frac {d^{2}}{d{x}^{2}}}r
 \left( x \right)-\\
\nonumber &-& 2\,{\frac {d}{dx}}U \left( x \right) -2\,A \left( x
 \right) U \left( x \right) r \left( x \right) {\frac {d^{3}}{d{x}^{3}
}}r \left( x \right) -2\, \left( {\frac {d}{dx}}U \left( x \right) 
 \right) r \left( x \right)  \left( {\frac {d}{dx}}r \left( x \right) 
 \right) {\frac {d}{dx}}A \left( x \right)+\\
\nonumber &+&2\,A \left( x \right) 
 \left( {\frac {d}{dx}}U \left( x \right)  \right)  \left( {\frac {d}{
dx}}r \left( x \right)  \right) ^{2}+ \left( {\frac {d}{dx}}U \left( x
 \right)  \right)  \left( r \left( x \right)  \right) ^{2}{\frac {d^{2
}}{d{x}^{2}}}A \left( x \right)-\\
\nonumber &-&2\,A \left( x \right)  \left( {\frac 
{d}{dx}}U \left( x \right)  \right) r \left( x \right) {\frac {d^{2}}{
d{x}^{2}}}r \left( x \right) + 2\,A \left( x \right) U \left( x
 \right)  \left( {\frac {d}{dx}}r \left( x \right)  \right) {\frac {d^
{2}}{d{x}^{2}}}r \left( x \right)  \Bigg) ,
\end{eqnarray}

\begin{eqnarray}
\nonumber C^{3}_{0}&=&-\frac{1}{4\left( r \left( x
 \right)  \right) ^{4}\sin \left( \theta \right)}\Bigg[A \left( x \right)  \left( {\frac {d}{d\theta}}Y \left( 
\theta \right)  \right)\Bigg] \Bigg( U \left( x \right)  \left( r \left( x
 \right)  \right) ^{2}{\frac {d^{3}}{d{x}^{3}}}A \left( x \right) -\nonumber\\ &-&
 4\,
U \left( x \right)  \left( {\frac {d}{dx}}A \left( x \right)  \right) 
r \left( x \right) {\frac {d^{2}}{d{x}^{2}}}r \left( x \right)-\\
\nonumber &-& 2\,{
\frac {d}{dx}}U \left( x \right) -2\,A \left( x \right) U \left( x
 \right) r \left( x \right) {\frac {d^{3}}{d{x}^{3}}}r \left( x
 \right) -2\, \left( {\frac {d}{dx}}U \left( x \right)  \right) r
 \left( x \right)  \left( {\frac {d}{dx}}r \left( x \right)  \right) {
\frac {d}{dx}}A \left( x \right) +\\
\nonumber &+& 2\,A \left( x \right)  \left( {
\frac {d}{dx}}U \left( x \right)  \right)  \left( {\frac {d}{dx}}r
 \left( x \right)  \right) ^{2}+ \left( {\frac {d}{dx}}U \left( x
 \right)  \right)  \left( r \left( x \right)  \right) ^{2}{\frac {d^{2
}}{d{x}^{2}}}A \left( x \right) -\\
&-& 2\,A \left( x \right)  \left( {\frac 
{d}{dx}}U \left( x \right)  \right) r \left( x \right) {\frac {d^{2}}{
d{x}^{2}}}r \left( x \right) +2\,A \left( x \right) U \left( x
 \right)  \left( {\frac {d}{dx}}r \left( x \right)  \right) {\frac {d^
{2}}{d{x}^{2}}}r \left( x \right)  \Bigg). \nonumber
\end{eqnarray}

The non-vanishing components of $T^{(\vartheta)\,\mu}_{\ \,\nu}$ are
\begin{eqnarray}
 T^{(\vartheta)\,0}_{\phantom{(\vartheta)}0}&=&-\frac{1}{2}\,{\frac { \left( r \left( x \right)  \right) ^{2}A \left( x
 \right)  \left( {\frac {d}{dx}}U \left( x \right)  \right) ^{2}
 \left( Y \left( \theta \right)  \right) ^{2}+ \left( U \left( x
 \right)  \right) ^{2} \left( {\frac {d}{d\theta}}Y \left( \theta
 \right)  \right) ^{2}}{ \left( r \left( x \right)  \right) ^{2}}},\\
T^{(\vartheta)\,1}_{\phantom{(\vartheta)}1}&=&\frac{1}{2}\,{\frac { \left( r \left( x \right)  \right) ^{2}A \left( x
 \right)  \left( {\frac {d}{dx}}U \left( x \right)  \right) ^{2}
 \left( Y \left( \theta \right)  \right) ^{2}- \left( U \left( x
 \right)  \right) ^{2} \left( {\frac {d}{d\theta}}Y \left( \theta
 \right)  \right) ^{2}}{ \left( r \left( x \right)  \right) ^{2}}},\\
T^{(\vartheta)\,1}_{\phantom{(\vartheta)}2}&=& A \left( x \right)  \left( {\frac {d}{dx}}U \left( x \right)  \right) 
Y \left( \theta \right) U \left( x \right) {\frac {d}{d\theta}}Y
 \left( \theta \right)\\
T^{(\vartheta)\,2}_{\phantom{(\vartheta)}1}&=& {\frac { \left( {\frac {d}{dx}}U \left( x \right)  \right) Y \left( 
\theta \right) U \left( x \right) {\frac {d}{d\theta}}Y \left( \theta
 \right) }{ \left( r \left( x \right)  \right) ^{2}}}\\
 T^{(\vartheta)\,2}_{\phantom{(\vartheta)}2}&=& -\frac{1}{2}\,{\frac { \left( r \left( x \right)  \right) ^{2}A \left( x
 \right)  \left( {\frac {d}{dx}}U \left( x \right)  \right) ^{2}
 \left( Y \left( \theta \right)  \right) ^{2}- \left( U \left( x
 \right)  \right) ^{2} \left( {\frac {d}{d\theta}}Y \left( \theta
 \right)  \right) ^{2}}{ \left( r \left( x \right)  \right) ^{2}}}\\
T^{(\vartheta)\,3}_{\phantom{(\vartheta)}3}&=& -\frac{1}{2}\,{\frac { \left( r \left( x \right)  \right) ^{2}A \left( x
 \right)  \left( {\frac {d}{dx}}U \left( x \right)  \right) ^{2}
 \left( Y \left( \theta \right)  \right) ^{2}+ \left( U \left( x
 \right)  \right) ^{2} \left( {\frac {d}{d\theta}}Y \left( \theta
 \right)  \right) ^{2}}{ \left( r \left( x \right)  \right) ^{2}}}.
\end{eqnarray}

  
\textbf{Acknowledgments.} Authors are grateful to G. Olmo for important discussions, and to K. Bronnikov and R. Konoplya for some criticism of the manuscript.
This work was partially supported by Conselho Nacional de Desenvolvimento Científico e Tecnológico (CNPq). The work by A. Yu. P. has been partially supported by the CNPq project No. 303783/2015-0.

\end{document}